# Shape of temperature dependence of spontaneous magnetization of various ferromagnets.

A. Perevertov[1*]

[1]Institute of Physics of the Czech Academy of Sciences, Department of Magnetic Measurements and Materials, 18200 Prague, Czech Republic

**ABSTRACT**. The shape of the temperature dependence of spontaneous magnetization was analyzed for about forty ferromagnetic materials. The squareness of the shape was determined by fitting the magnetization curves with the superellipse equation (Lamé curve). The agreement between the Lamé curve fits and the experimental data was good for most materials. The squareness parameter (the power coefficient in the superellipse equation was found to range from 1.4 to 3.0. The largest squareness was observed for iron, whereas the smallest was found for antiferromagnetic materials and the $Ni_{55}Cu_{45}$ alloy. The squareness parameter was studied as a function of the Curie temperature, $T_C$. For metallic alloys, a general tendency was observed: the squareness increased with increasing Curie temperature. The only exception was cobalt, which exhibited the same magnetization curve in reduced coordinates as nickel despite having a Curie temperature twice as high. The addition of either ferromagnetic or nonferromagnetic metals to iron or nickel led to a decrease in squareness. No influence of the thermal expansion coefficient on the magnetization curve was observed: zero-expansion Invar exhibited a standard shape following the Lamé curve.

## 1. Introduction

Spontaneous magnetization of ferromagnetic materials, $M_S$ is strongly influenced by temperature, $T$ – $M_S$ decreases with temperature increase and vanishes above the Curie temperature, $T_C$. Understanding the magnetization curve, $M_S(T)$ is of great importance not only from fundamental but also from application points of view – design of magnetic material with certain properties in a given temperature range, spintronics and data processing, spin caloritronics, magnetocalorics, engineering of zero-expansion materials and so on [1-15].

The first model of the connection between magnetization and temperature was proposed by Weiss in 1907 [16-18]. It is based on the assumption that atomic vibrations disorient magnetic moments of electrons that are ordered due to a Weiss molecular field, which is proportional to the magnetization. The coupling between atomic vibrations and spins was introduced by making equal the kinetic energy of atoms and the magnetostatic energy of magnetic moments in a Weiss molecular field. The theory of ferromagnetism by Weiss is based on the paramagnetism model of Langevin. Later it was extended to the more general Brillouin theory of localized magnetic moments [16-18].

In the classical thermodynamics vibrations and magnetic moments are decoupled – vibration of a body does not produce its rotation. Vibrations can't make a magnetic moment rotate by 90° or 180° in the classical electrodynamics. Such a coupling can be introduced only using quantum mechanics principles.

There are several mechanisms of the influence of nuclei vibrations (temperature) on magnetic moment of electrons - spin–lattice coupling, exchange interaction modulation, and spin–orbit coupling effects [5-7]. Developing quantum mechanics based models of magnetization as a function of temperature is still a hot topic for research [4-13]. The problem is so complex that even prediction of a single parameter – the Curie temperature is a challenging task. In contrast to the Curie temperature, the shape of the magnetization curve as a function of temperature have never been compared for different materials. In all classical textbooks only the shape of the magnetization curve for nickel is discussed [16-18]. The problem here is that it is hardly possible to compare all curves as datasets – one needs a single parameter describing the curve shape. In the classical Brillouin theory the curve shape is indeed described by a single parameter – the angular momentum, $J$. Unfortunately, all possible Brillouin functions cover a small area of possible values – curves are quite close to each other. As we will show below, most of experimental curves for different materials are outside of these area, including even iron, nickel and cobalt. In 2005 Kuz'min addressed this problem and proposed a simple empirical formula with just two parameters. He calculated these parameters for several ferromagnetic materials [19].

## 2. Methods.

Recently we have shown that the spontaneous magnetization dependence on temperature in the reduced coordinates obeys the superellipse equation (Lamé curve) for N2MnGa alloy, iron, cobalt, nickel and gadolinium [20]:

$$(M_S/M_0)^{\eta} + (T/T_C)^{\eta} = m^{\eta} + \tau^{\eta} = 1 \quad (1)$$

where $M_S$ is the spontaneous magnetization at temperature $T$; $M_0 = M_S(T=0)$; $T_C$ – the Curie temperature, $m$ - reduced magnetization and $\tau$ is reduced temperature.

Here we have a single parameter, $\eta$ describing squareness of the $m(\tau)$ curve. As $\eta$ approaches infinity, the curve approaches the perfect square shape – temperature has no effect on the magnetization till $T_C$, at which magnetization drops abruptly to zero. At $\eta = 1$ the magnetization changes at the same rate starting already from zero temperature. So, $\eta$ should reflect coupling strength between atomic vibrations (temperature) and the magnetic moments of electrons.

In this work we analyze experimental $m(\tau)$ data for different alloys using the superellipse equation and determine the power coefficient $\eta$ (squareness, critical exponent) of $m(\tau)$ curve for different materials based on experimental data present in the literature. Data was extracted by manual digitalization of curves from articles figures using Corel Technical Designer software except for iron, nickel and cobalt the data, for which data was published as datasets [21].

## 3. Results

In Fig. 1 $m(\tau)$ curves are shown for four basic ferromagnetic elements – Fe, Ni, Co [21] and Gd [22].

The experimental data follows the Lamé curve with good accuracy. The power coefficient is largest for iron ($\eta = 3.0$) and smallest for gadolinium ($\eta = 2.05$). For nickel and cobalt the power coefficient is the same – 2.7.

In Fig. 2 the experimental data are shown for Fe-Ni [23-24], Ni-Co [25] and $Fe_{40}Co_{60}$ [17] alloys. The power coefficient is smallest for invar ($\eta = 1.6$) [23-24]. Surprisingly, the power coefficient for Ni-Co alloys is the same for all compositions including pure nickel and cobalt. The power coefficient 2.5 is lower than that for cobalt and nickel from the data by other authors [21], which was 2.7 (see Fig. 1). Ni-Co measurements were done in 1910 and the measurement error is several times larger comparing to more recent measurements – data points lie between Lamé curves from $\eta = 2.4$ to $\eta = 2.7$.

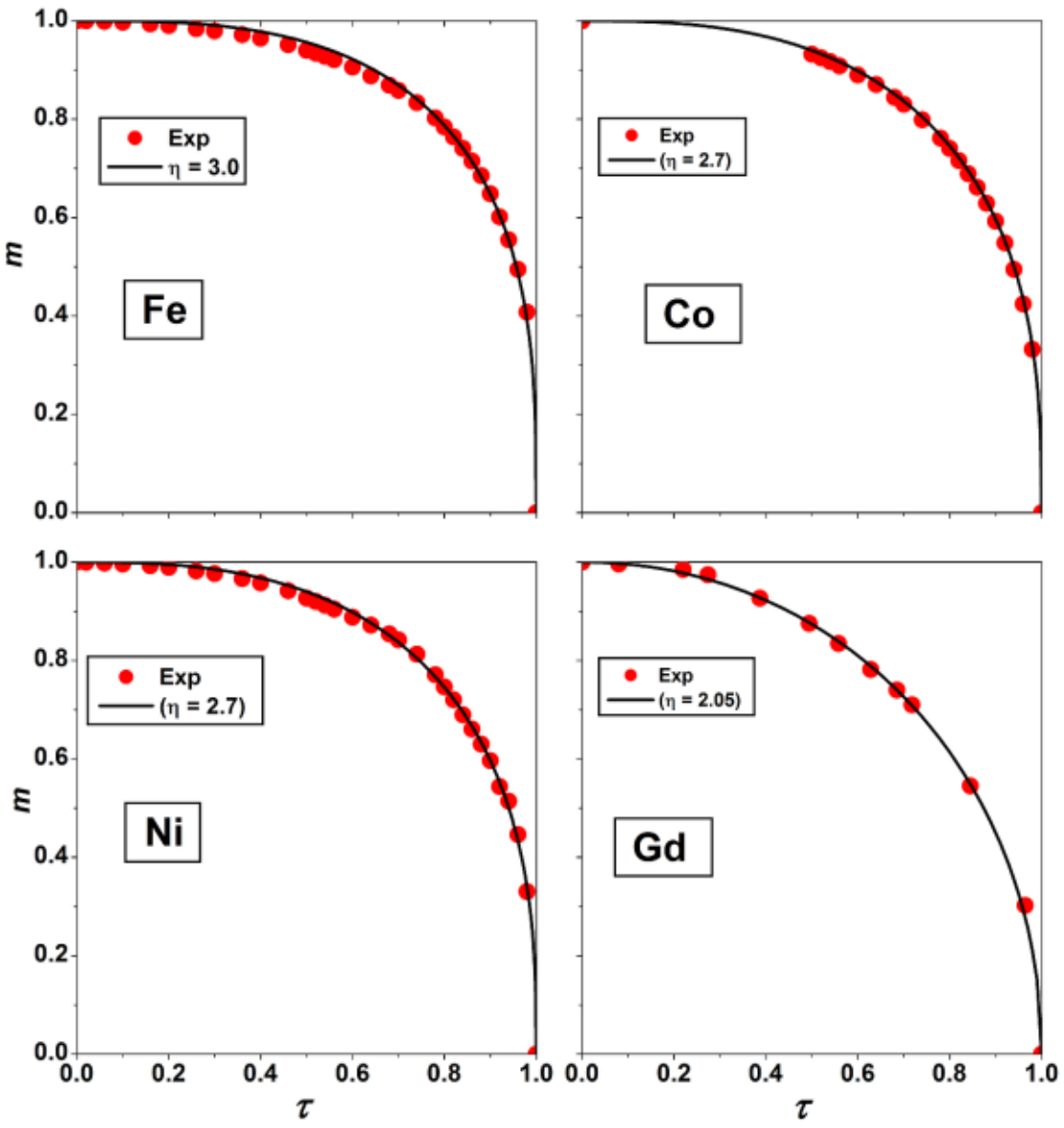

Fig. 1. $m(\tau)$ curves for Fe, Ni, Co (data taken from ref. [21]) and Gd (data taken from ref. [22]) and corresponding fits by the superellipse equation.

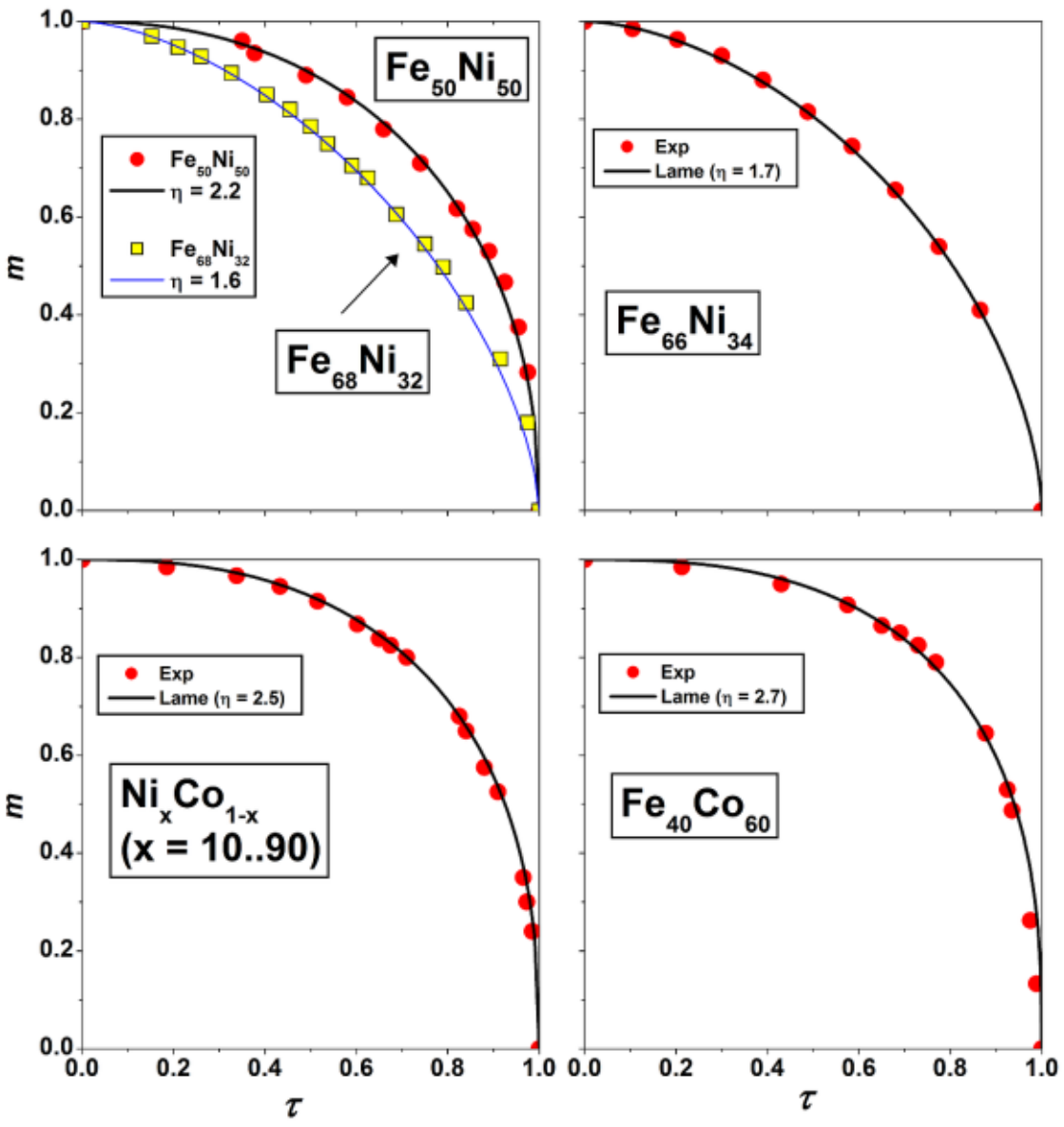

Fig. 2. $m(\tau)$ curves for Fe-Ni (data taken from refs. [23-24]), Ni-Co (data taken from ref. [25]) and $Fe_{40}Co_{60}$ (data taken from ref. [17]) alloys.

In Fig. 3 the experimental data and corresponding fits by the superellipse equation are shown for Ni-Cu alloys [26]. The power coefficient decreases with increase of copper content down to $\eta = 1.5$ for $Ni_{55}Cu_{45}$.

In Fig. 4 data for $XCo_5$ alloys ($YCo_5$, $NdCo_5$, $CeCo_5$, $PrCo_5$ and $CeMMCo_5$,)) is shown [27].

In Fig. 5 data for Gd-Y alloys is shown [28]. Here the experimental data deviates from a symmetrical Lamé curve.

In Fig.6. $m(\tau)$ curves for EuO [29], BaTiO3 [30], $Gd_{90}La_{10}$ [28] and EuS [31] are shown. All data except for $Gd_{90}La_{10}$ is well described by the Lamé curve. The data for $Gd_{90}La_{10}$ was taken from the same work as for Gd-Y alloys [28], there the experimental curves also showed a little asymmetry (see Fig.5).

In Fig.7. data for FeCo-V alloys [32], $Ni_{49}Mn_{21}Ga_{26}Cu_4$ (our measurement), $Ni_{50}Mn_{24}Ga_{26}$ (our measurement) and antiferromagnets ($K_2Fe_{1-x}In_xCl_5H_2O$, $Rb_2Fe_{1-x}In_xCl_5H_2O$ and $Mn_{1-x}Zn_xF_2$) [33] is presented. All experimental curves are in a good agreement with the superellipse equation.

And finally, in Fig. 8. $m(\tau)$ curves for $Co_{89}Cu_{11}$ [34], $Fe_{79}Si_9B_{13}$ [35], $\alpha$-$Fe_{80}B_{20}$ [36] and $\alpha$-$Fe_2O_3$ [37] are shown.

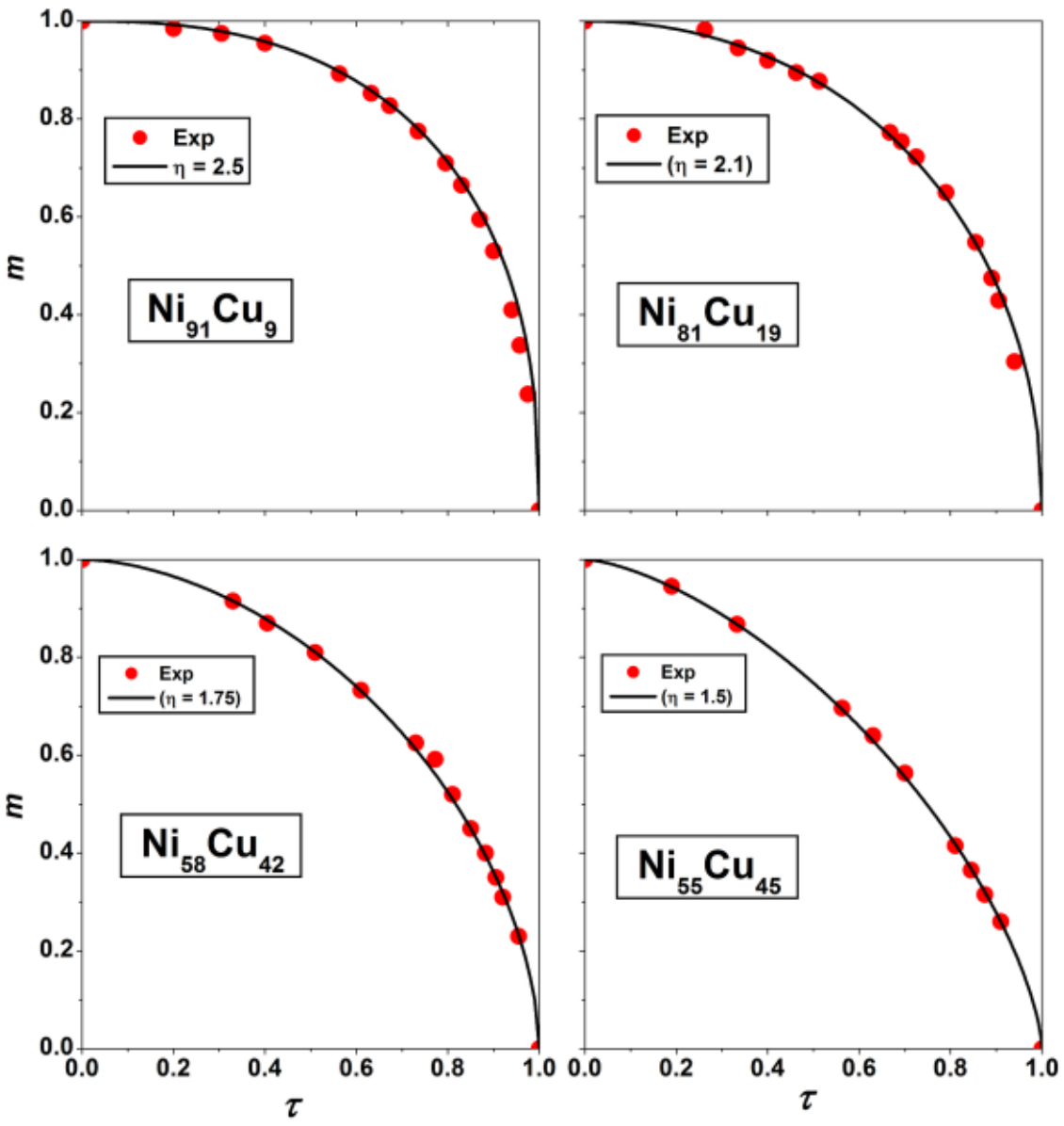

Fig. 3. $m(\tau)$ curves for Ni-Cu alloys (data taken from ref. [26]).

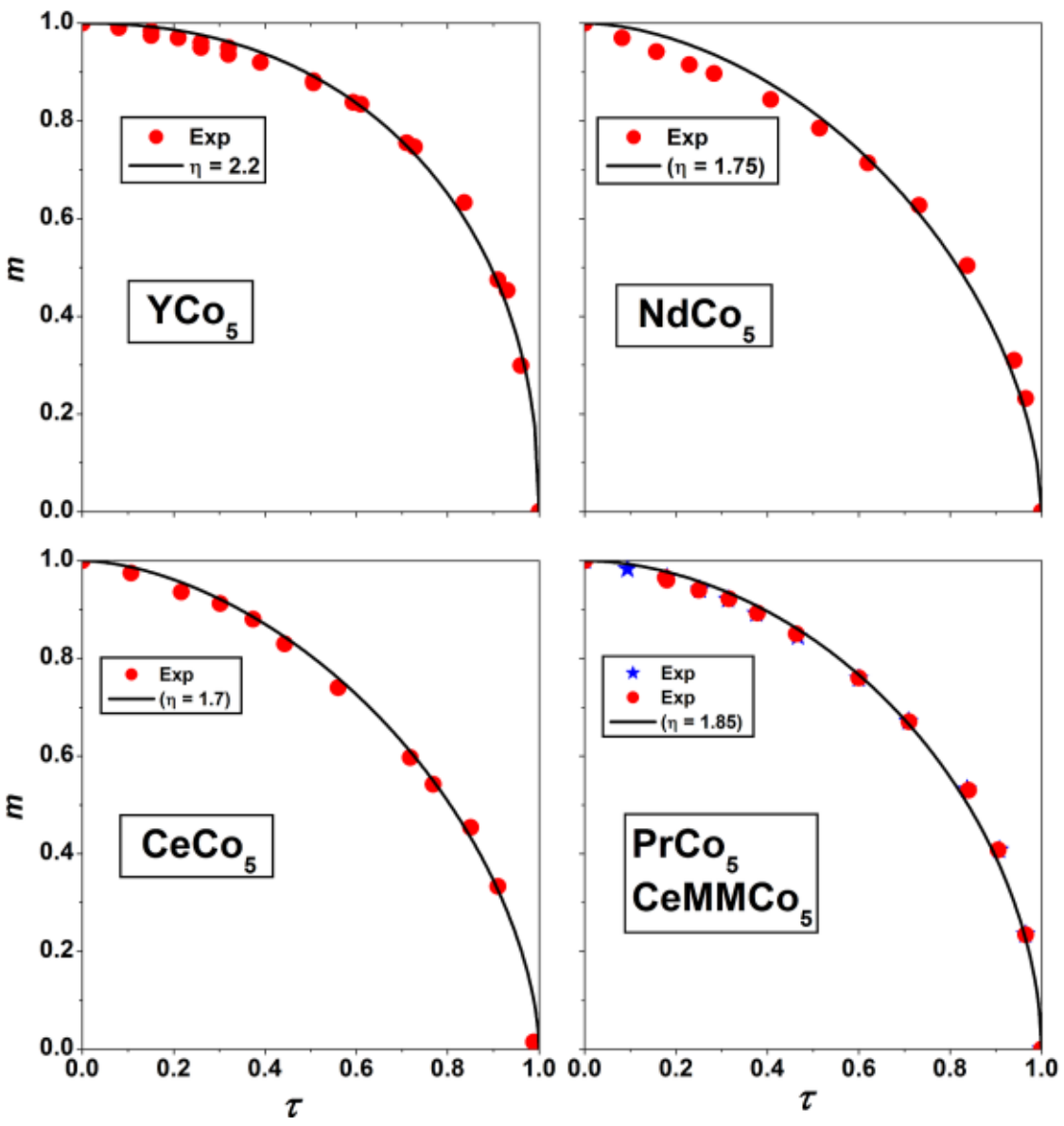

Fig. 4. *m( τ)* curves for $XCo_5$ alloys ($YCo_5$, $NdCo_5$, $CeCo_5$, $PrCo_5$ and $CeMMCo_5$,)) (data taken from ref. [27]).

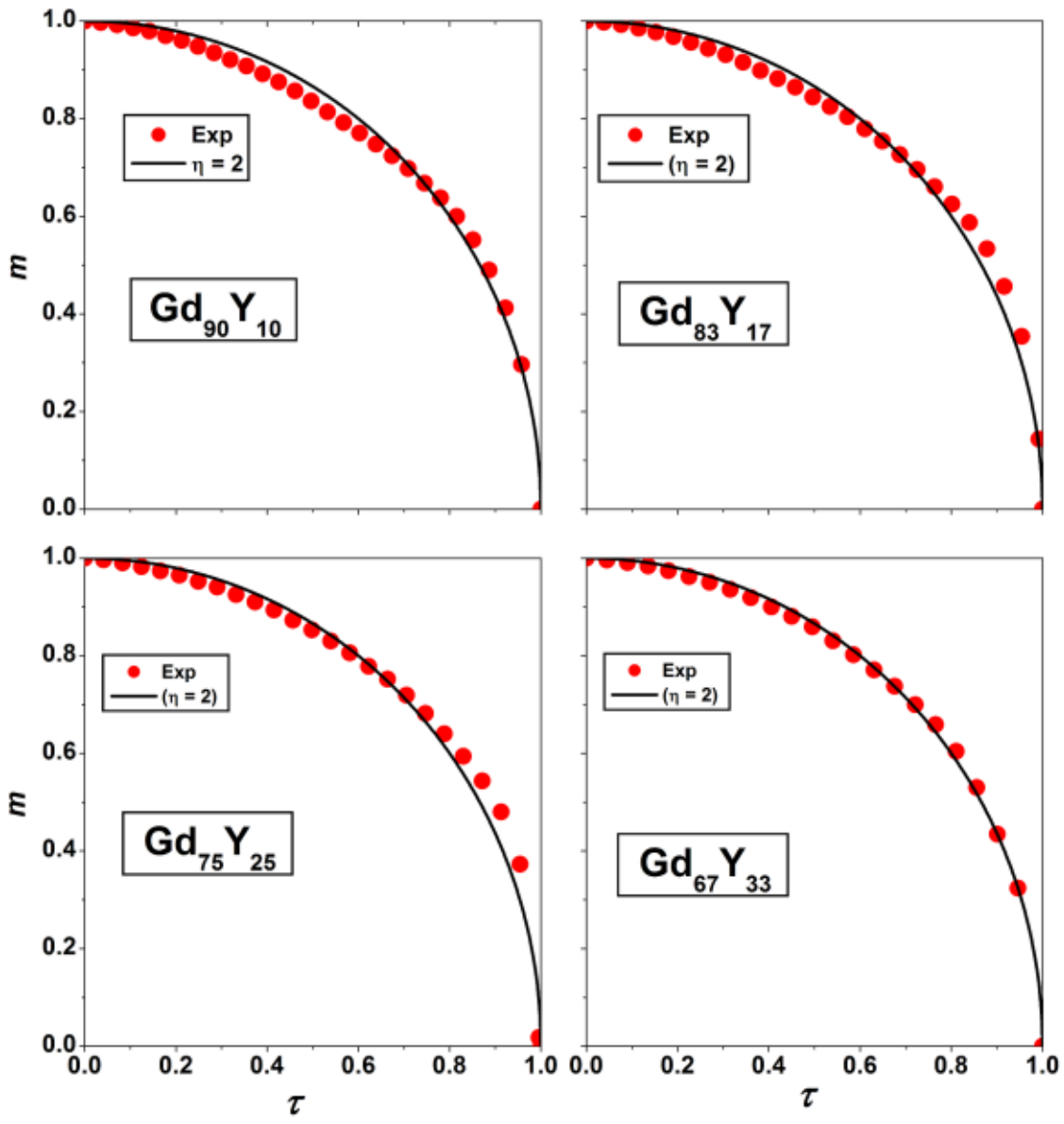

Fig. 5. *m( τ)* curves for Gd-Y alloys (data taken from ref. [28]).

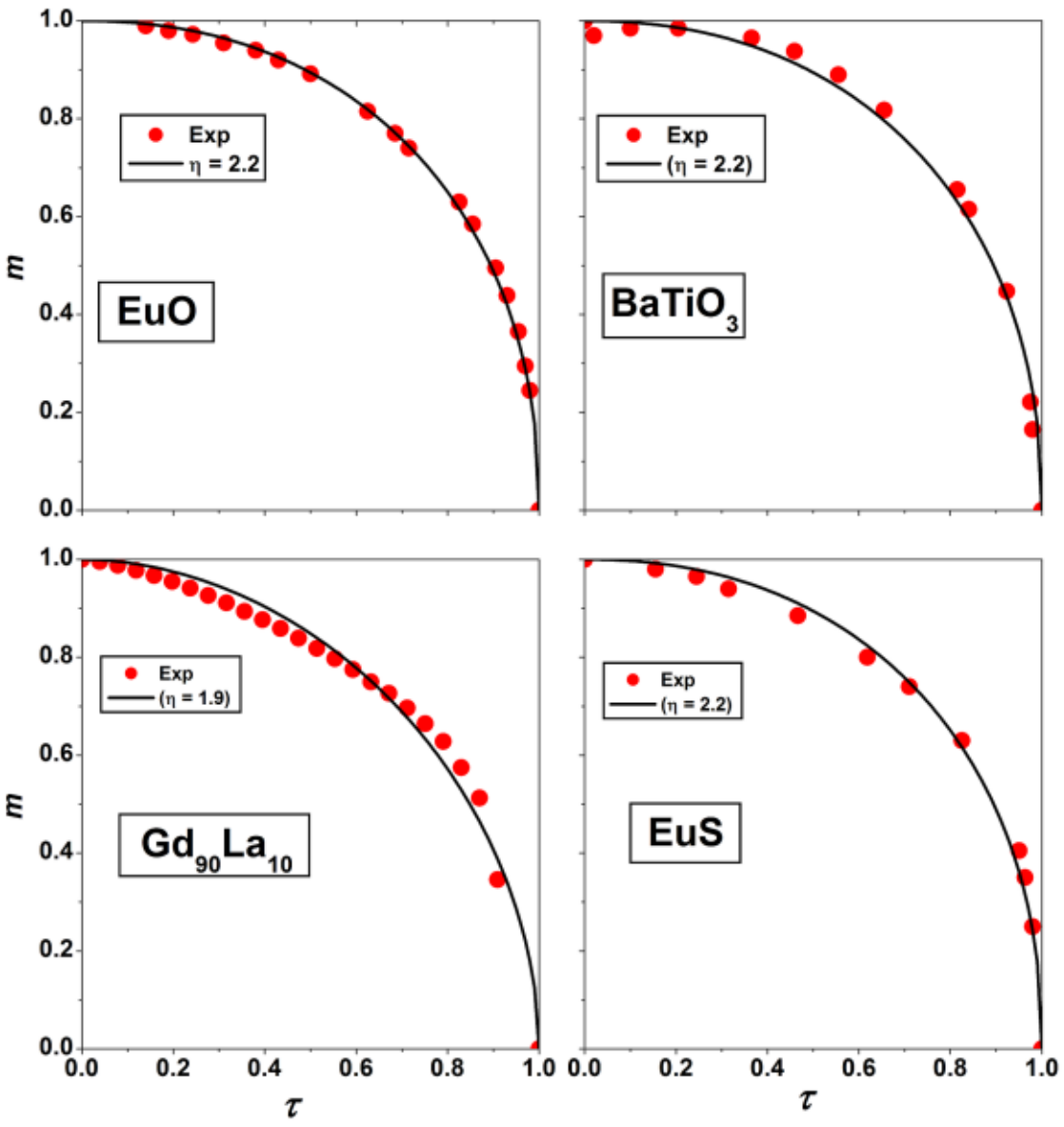

Fig. 6. *m( τ)* curves for EuO (data taken from ref. [29]), BaTiO3 [30], $Gd_{90}La_{10}$ (data taken from ref. [28]) and EuS (data taken from ref. [31]).

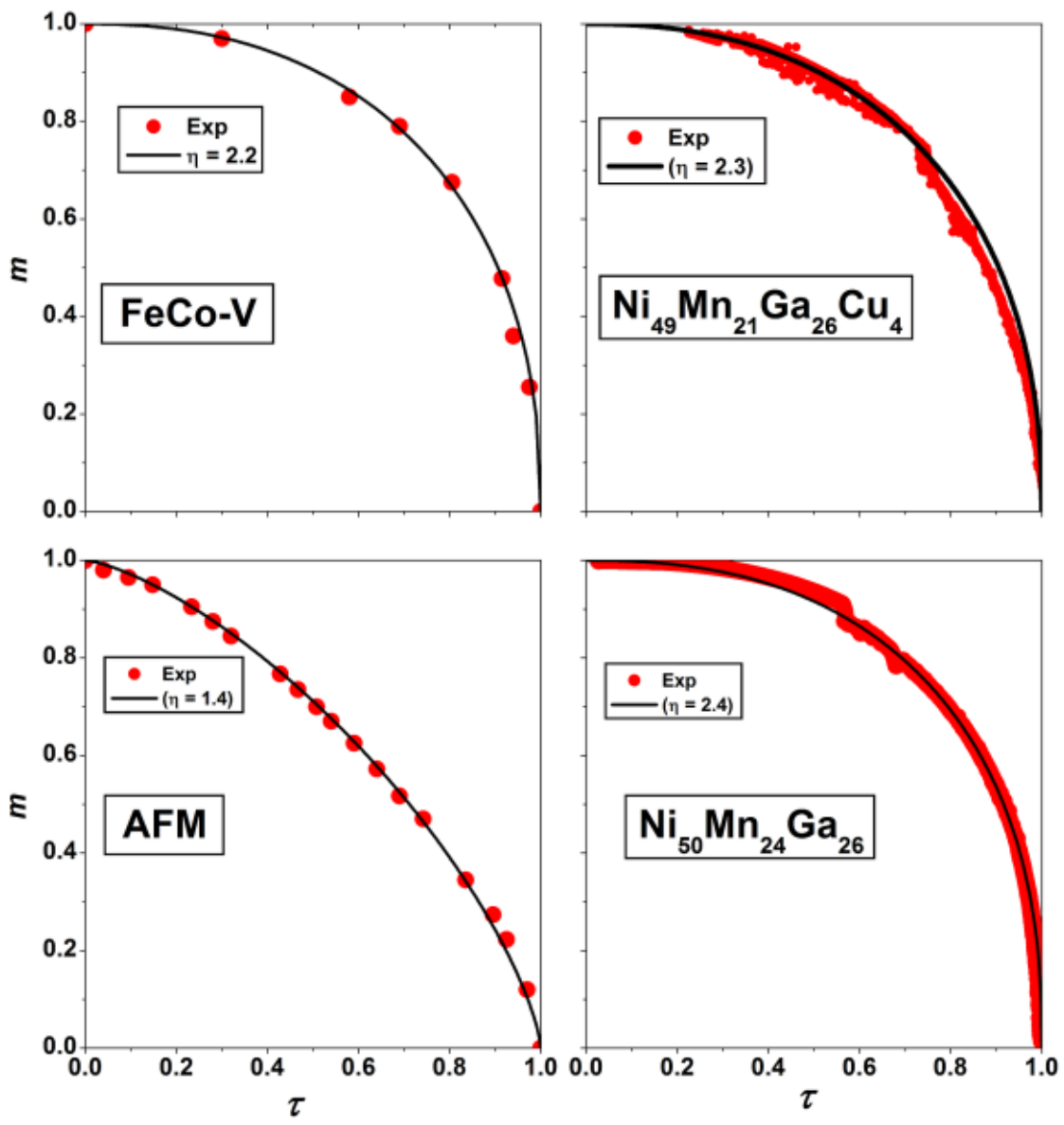

Fig. 7. *m(τ)* curves for FeCo-V alloys (data taken from ref. [32]), $Ni_{49}Mn_{21}Ga_{26}Cu_4$ (our measurement), $Ni_{50}Mn_{24}Ga_{26}$ (our measurement) and antiferromagnets ($K_2Fe_{1-x}In_xCl_5H_2O$, $Rb_2Fe_{1-x}In_xCl_5H_2O$ and $Mn_{1-x}Zn_xF_2$) (data taken from ref. [33]).

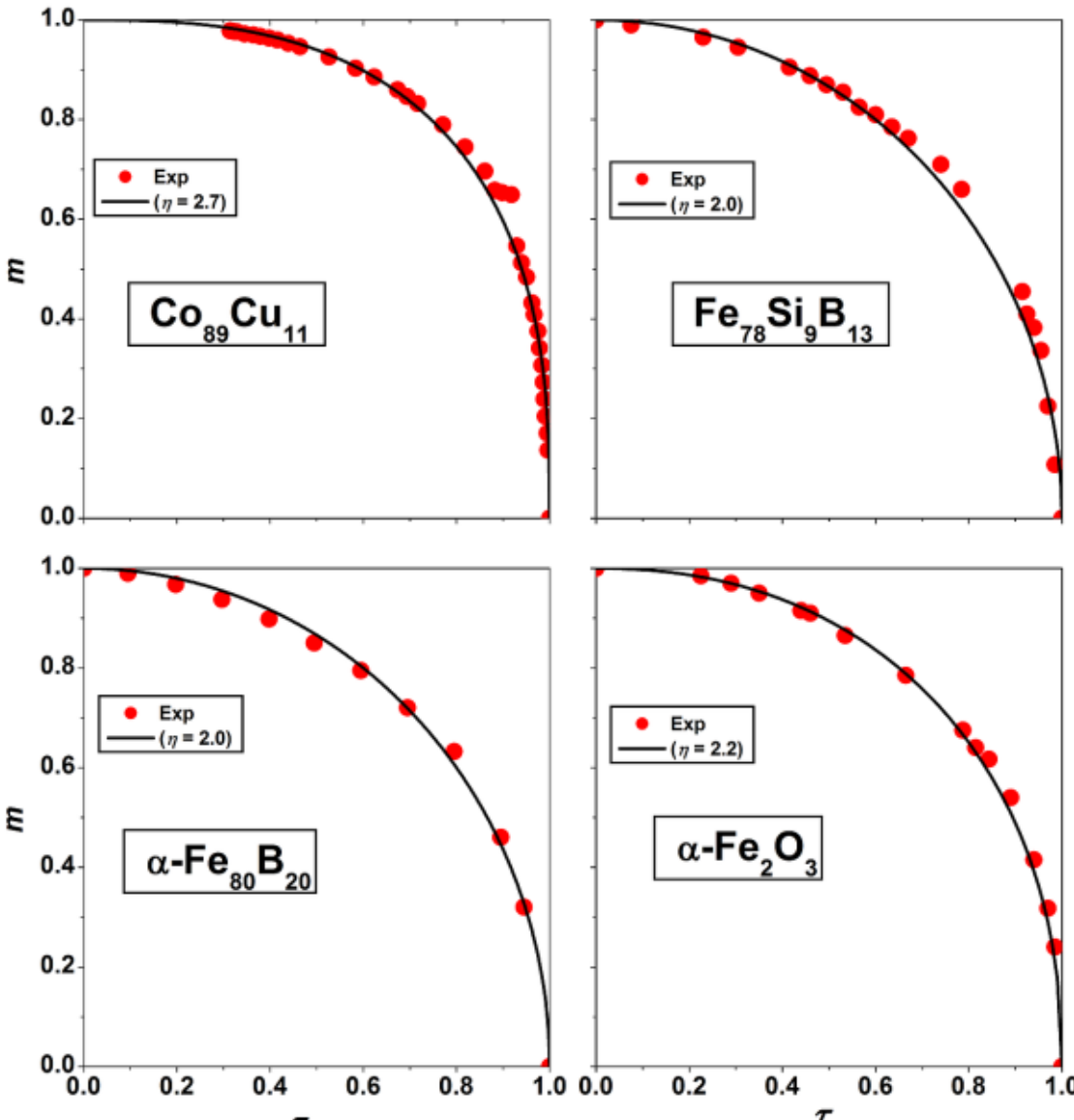

Fig. 8. *m(τ)* curves for $Co_{89}Cu_{11}$ (data taken from ref. [34]), $Fe_{79}Si_9B_{13}$ (data taken from ref. [35]), α-$Fe_{80}B_{20}$ (data taken from ref. [36]) and α-$Fe_2O_3$ (data taken from ref. [37]).

Above we have shown that the superellipse equation agrees well with experimental data on about forty materials including antiferromagnets. The only exception are Gd-Y alloys, where the experimental *m(τ)* data deviates from the symmetrical Lamé curve (see Fig. 5). Since all measurements were done in one work [28], one can't tell if it was a measurement artifact or a real property of materials. Asymmetrical *m(τ)* curves can be described by a modified Eq. (1) with different power coefficients for magnetization and temperature.

$$m^a + \tau^b = 1 \qquad (2)$$

In Fig. 9 the magnetization curves of Gd-Y alloys [28] are fitted by this modified Eq. (2).

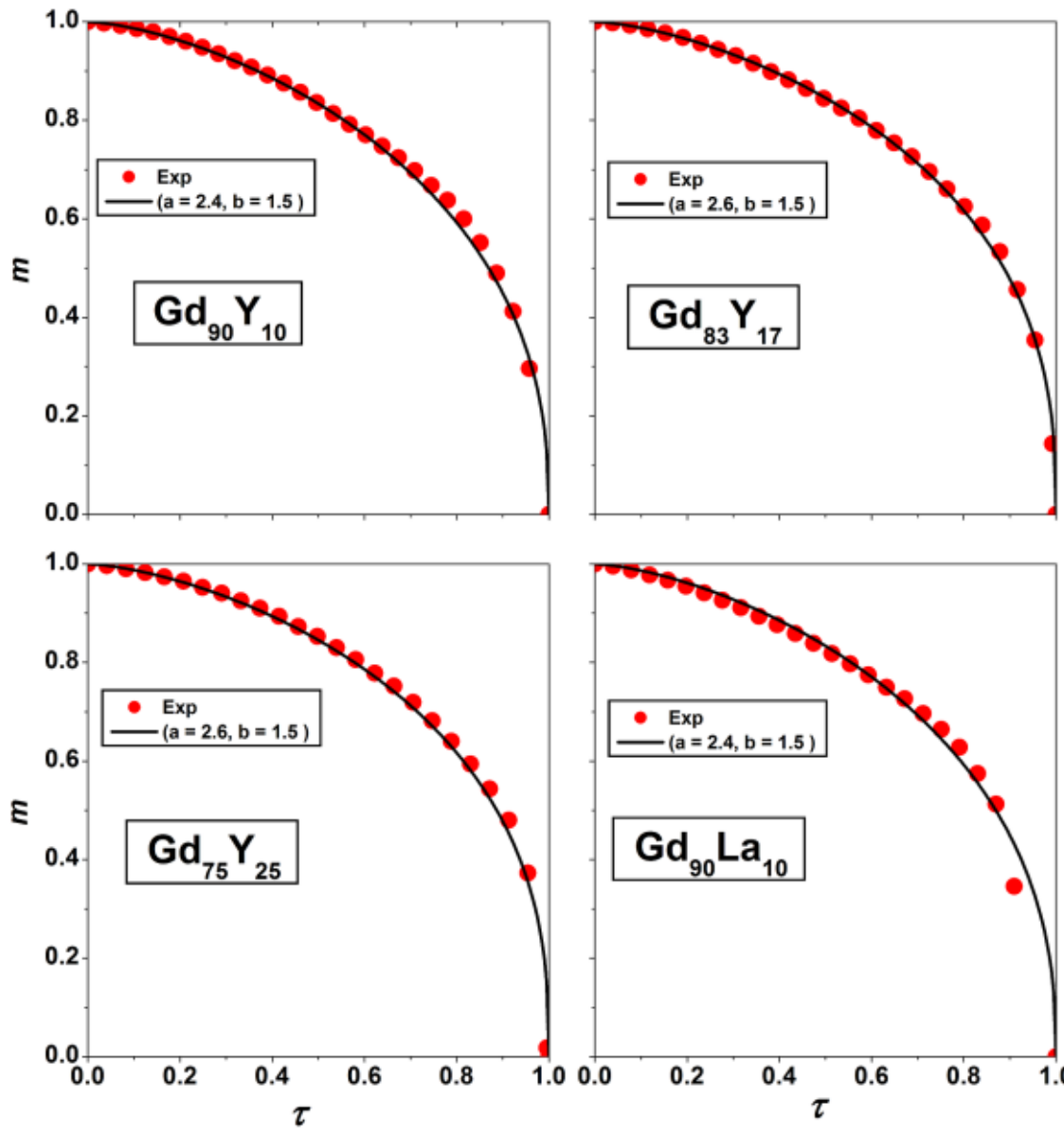

Fig. 9. $m(\tau)$ curves for Gd-Y alloys (data taken from ref. [28]) with corresponding fits using the Eq. (2).

## 4. Discussion

The power coefficient, $\eta$ is the parameter that characterizes the temperature dependence of the magnetization of a material in addition to the Curie temperature $T_C$. It gives squareness of the $m(\tau)$ curve. The larger $\eta$ – the squarer is the curve. Maximum squareness shows iron ($\eta = 3.0$), the smallest – antiferromagnetic materials ($\eta = 1.4$) (see Fig. 7) [33]. A reasonable question arises: is there a connection between these parameters. In Fig. 10 the power coefficient, $\eta$ is plotted as a function of $T_C$ for all materials we analyzed in this work. For metallic materials one can see the tendency - $\eta$ increases with $T_C$ increase (the blue dashed curve in Fig. 10). It means that the squareness of curve increases with $Tc$ increase. This tendency holds till iron. The squareness of the cobalt curve is the same as that of nickel ($\eta = 2.7$) despite of much higher $T_C$. In fact, the $m(\tau)$ curves for cobalt and nickel taken from the data in ref. [21] are nearly identical. For X-$Co_5$ materials and oxides the parameter is lower. The same power coefficient for all Ni-Co alloys for fata from [25] looks very interesting.

The experimental $m(\tau)$ data analyzed in this work was obtained by different authors at different time on different equipment using different techniques. So, it is natural to expect that some data may contain measurements errors. Nevertheless, we see that the superellipse equation gives good agreement with most of experimental data. The error of $\eta$ determination by fitting experimental data by Lamé curve using minimization of mean squared error is about ±0.025 (see supplementary Fig. 1). Measurement error of magnetization in VSM is usually about 2% and the digitalization error is about 1%. So, altogether we expect the error of $\eta$ determination for most materials around ±0.05. The exception is Ni-Co alloys, where the measurement scatter is several times larger (see ref. 25) and we estimate the error three times larger – about ±0.15.

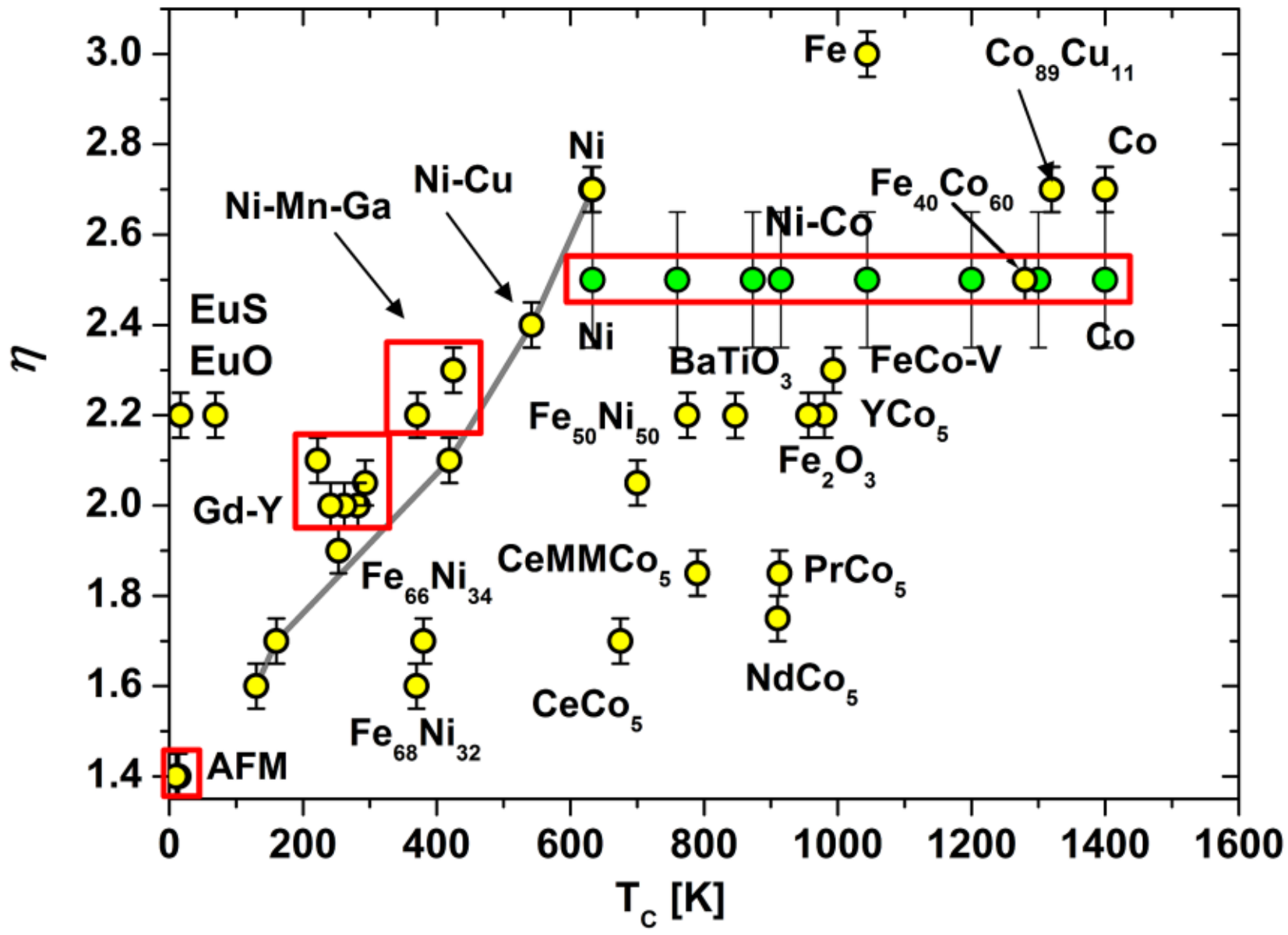

Fig. 10. The squareness parameter, $\eta$ as a function of the Curie temperature, $T_C$ for different materials. Data points for Ni-Cu alloys are connected by a gray line. Data point for Ni-Co alloys are shown as green circles.

Looking at data in Fig. 10 one can guess that the lower limit for the squareness parameter should be around 1.35. The maximum power coefficient, $\eta = 3.0$ was observed for iron, which stays above all other materials.

In the Brillouin theory spontaneous magnetization in the absence of the applied field is given by:

$$\frac{M_S}{M_0} = B_J\left(\frac{3J}{J+1}\frac{M_S}{M_0}\frac{T_C}{T}\right), \tag{3}$$

where $J$ is the total angular momentum. $B_J$ – the Brillouin function.

The squareness of the magnetization curve is defined by the total angular momentum, $J$. The lower $J$ – the larger is squareness. $J$ is 4 for iron and nickel, 9/2 for cobalt and 7/2 for gadolinium. The experimental $m(\tau)$ curve for nickel is far from the Brillouin curve for $J = 4$ and is closer to that for $J = 1/2$. The curve is a little outside the Brillouin function [18]. The experimental $m(\tau)$ curve for iron is even more square than the nickel curve. From other side, in the Brillouin theory $J$ cannot be lower than 1/2 for which the curve squareness is largest. For gadolinium the closest approximation by the Brillouin function is for $J$ close to infinity.

In Gadolinium magnetism comes from localized 4f electrons. Each Gd atom has a large moment (~7 μB), but the 4f electrons are very tightly localized and do not overlap between atoms. Therefore, spins do not interact directly, they couple via indirect exchange (RKKY interaction), mediated by conduction electrons – relatively weak coupling [18]. In cobalt, nickel and iron, from other side, magnetism originates from itinerant 3d electrons. 3d orbitals extend and overlap strongly between atoms. This allows strong direct exchange interaction. As the result, we observe much higher squareness of $m(t)$ curves for gadolinium than for iron, nickel and cobalt. To clarify the influence of the localization of electrons on the $m(\tau)$ squareness additional experiments and investigations are necessary. The most promising direction is to study Ni-Cu alloys, for which the squareness parameter, $\eta$ decreases from 2.7 to 1.5 with increase of cupper content (see Fig. 3) thus covering nearly all range of observed values of $\eta$. In Co-Cu alloy, from other side, increase of copper content up to 11% has very little effect on $T_C$ and has no effect on the squareness ( $\eta = 2.7$ ) [34].

It is very interesting that all Ni-Co alloys have the same $m(\tau)$ shape despite of TC change from 630 K for pure nickel to 1400 K for pure nickel. The Curie temperature is set by the absolute strength of the exchange interactions between spins. Cobalt has stronger exchange interactions and larger exchange splitting of the 3d bands comparing to nickel [18]. As the result, cobalt has higher $T_C$ (1400 K) than nickel (630 K). From other side, Co and Ni belong to the same universality class (3D Heisenberg-like ferromagnets) with similar spin interactions, so their reduced magnetization curves coincide. It was pointed out [25] that Ni-Co alloys is the only exception among all nickel alloys – the shape of the $m(\tau)$ curve for of all other alloys strongly changes with the reduction of nickel content. Cobalt is in the low-temperature hcp phase below 700 K and in the high-temperature fcc phase above 700 K. Surprisingly, this phase transition has no visible effect on the m(t) curve despite of large lattice distortion – $c/a = 1.62$ in the hcp phase.

In the classical mechanics a body vibration does not produce its rotation. It can just increase rotation in a resonant system. In the Brillouin theory the kinetic energy of atomic nuclei is compared with the energy of magnetic moments oriented at an angle to the (internal) magnetic field. The mechanism of coupling between nuclei vibrations and carriers of magnetic moments (electrons) is not proposed. Obviously, if that coupling is lower, the temperature has lower effect on the spins orientation – the Curie temperature and squareness of the magnetization curve should both increase. If magnetic moments are strongly coupled to nuclei, atomic vibrations have stronger effect on magnetic moments – $T_C$ and $\eta$ should both decrease. Following this hypothesis, $\eta$ and $T_C$ in amorphous materials should be lower comparing to crystalline materials. Data for $Fe_{80}B_{20}$ and $Fe_{78}Si_9B_{13}$ alloys [35-36] indeed yields low $\eta$ = 2.0 as well as for $\alpha$-$Fe_2O_3$ (see Fig. 7) [37]. Alloying ferromagnetic nickel with ferromagnetic iron or with nonmagnetic copper has the same effect on the $m(\tau)$ curve – the squareness gradually decreases (see Figs. 2,3 and 10). The $m(\tau)$ curves of invar and $Ni_{55}Cu_{45}$ alloys are nearly identical. Invar is known for nearly zero expansion coefficient in a wide range of temperature. From Fig. 10 we see that it is also lies much lower than other metallic alloys – $\eta$ = 1.6 instead of expected value of 2.

All values of $T_C$ and $\eta$ in Fig. 10 are presented in the Table I together with the data source references and the coefficient of determination, $R^2$.

Table I. $T_C$ and $\eta$ values for different materials.

| Material | $T_C$ ($T_N$) [K] | $\eta$ | Ref. | $R^2$ | Fig. # |
|---|---|---|---|---|---|
| Fe | 1044 | 3.0 | 21 | 0.9978 | 1 |
| Co | 1400 | 2.7 | 21 | 0.9993 | 1 |
| Ni | 631 | 2.7 | 21 | 0.9988 | 1 |
| Gd | 293 | 2.05 | 22 | 0.9993 | 1 |
| $Fe_{50}Ni_{50}$ | 775 | 2.2 | 23 | 0.9969 | 2 |
| $Fe_{66}Ni_{34}$ | 380 | 1.7 | 24 | 0.9998 | 2 |
| $Fe_{68}Ni_{32}$ | 370 | 1.6 | 23 | 0.9975 | 2 |
| $Fe_{40}Co_{60}$ | 1283 | 2.5 | 17 | 0.9829 | 2 |
| Ni | 633 | 2.5 | 25 | 0.9975 | 2 |
| $Ni_{90}Co_{10}$ | 760 | 2.5 | 25 | 0.9975 | 2 |
| $Ni_{80}Co_{20}$ | 873 | 2.5 | 25 | 0.9975 | 2 |
| $Ni_{75}Co_{25}$ | 915 | 2.5 | 25 | 0.9975 | 2 |
| $Ni_{60}Co_{40}$ | 1044 | 2.5 | 25 | 0.9975 | 2 |
| $Ni_{40}Co_{60}$ | 1200 | 2.5 | 25 | 0.9975 | 2 |
| $Ni_{20}Co_{80}$ | 1300 | 2.5 | 25 | 0.9975 | 2 |
| Co | 1400 | 2.5 | 25 | 0.9975 | 2 |
| $Ni_{91}Cu_9$ | 542 | 2.4 | 26 | 0.9940 | 3 |
| $Ni_{81}Cu_{19}$ | 419 | 2.1 | 26 | 0.9955 | 3 |
| $Ni_{58}Cu_{42}$ | 160 | 1.7 | 26 | 0.9987 | 3 |
| $Ni_{55}Cu_{45}$ | 130 | 1.6 | 26 | 0.9998 | 3 |
| YCo5 | 980 | 2.2 | 27 | 0.9960 | 4 |
| NdCo5 | 910 | 1.75 | 27 | 0.9916 | 4 |
| CeCo5 | 675 | 1.7 | 27 | 0.9935 | 4 |
| PrCo5 | 913 | 1.85 | 27 | 0.9975 | 4 |
| CeMMCo5 | 790 | 1.85 | 27 | 0.9975 | 4 |
| $Gd_{90}Y_{10}$ | 282 | 2.0 | 28 | 0.9840 | 5 |
| $Gd_{83}Y_{17}$ | 262 | 2.0 | 28 | 0.9905 | 5 |
| $Gd_{75}Y_{25}$ | 241 | 2.0 | 28 | 0.9890 | 5 |
| $Gd_{67}Y_{33}$ | 222 | 2.1 | 28 | 0.9940 | 5 |
| $Gd_{90}La_{10}$ | 253 | 1.9 | 28 | 0.9900 | 6 |
| $BaTiO_3$ | 847 | 2.2 | 30 | 0.9925 | 6 |
| EuO | 69 | 2.2 | 29 | 0.9988 | 6 |
| EuS | 16.6 | 2.2 | 31 | 0.9956 | 6 |
| FeCo-V | 993 | 2.3 | 32 | 0.9967 | 7 |
| $Ni_{50}Mn_{24}Ga_{26}$ | 371 | 2.4 | 20 | 0.9936 | 7 |
| $Ni_{49}Mn_{21}Ga_{26}Cu_4$ | 425 | 2.2 | -- | 0.9898 | 7 |
| $Mn_{1-x}Zn_xF_2$ | -- | 1.4 | 33 | 0.9989 | 7 |
| $K_2Fe_{1-x}In_xCl_5H_2O$ | 14 | 1.4 | 33 | 0.9989 | 7 |
| $Rb_2Fe_{1-x}In_xCl_5H_2O$ | 10 | 1.4 | 33 | 0.9989 | 7 |
| $Co_{89}Cu_{11}$ | 1320 | 2.7 | 34 | 0.9928 | 8 |
| $Fe_{78}Si_9B_{13}$ | 700 | 2.05 | 35 | 0.9938 | 8 |
| $Fe_{80}B_{20}$ | 647 | 2.0 | 36 | 0.9979 | 8 |
| $\alpha$-$Fe_2O_3$ | 956 | 2.2 | 37 | 0.9969 | 8 |

In the superellipse equation the squareness of the curve is defined by the power coefficient, $\eta$. Brillouin curves for different $J$ can be approximated by the Lamé function with different $\eta$. The agreement is not very good since for the Lamé functions $m(\tau)=\tau(m)$, and the Brillouin curve is always asymmetrical. However, the Brillouin curves can be approximated with a good accuracy by the modified Eq. (2) [20]. The results of the approximation of the Brillouin curves by Eqs. (1) and (2) are given in the table 2. We see that the Brillion theory can approximate only curves with $\eta$ from 2 to 2.6, while experimental data are in a wider range with $\eta$ from 1.4 to 3.0 (see Figs. 10 and 11).

Table 2. Fits coefficients by the superellipse Eq. (1) and by more general Eq. (2) of Brillouin functions for different $J$.

| | $J = 1/2$ | $J = 1$ | $J = 2$ | $J = 5/2$ | $J = 7/2$ | $J = \infty$ |
|---|---|---|---|---|---|---|
| $\eta$ | 2.6 | 2.4 | 2.3 | 2.2 | 2.1 | 2.0 |
| $a$ | 1.9 | 1.7 | 1.8 | 1.8 | 1.9 | 1.3 |
| $b$ | 3.4 | 3.2 | 2.6 | 2.5 | 2.2 | 2.2 |

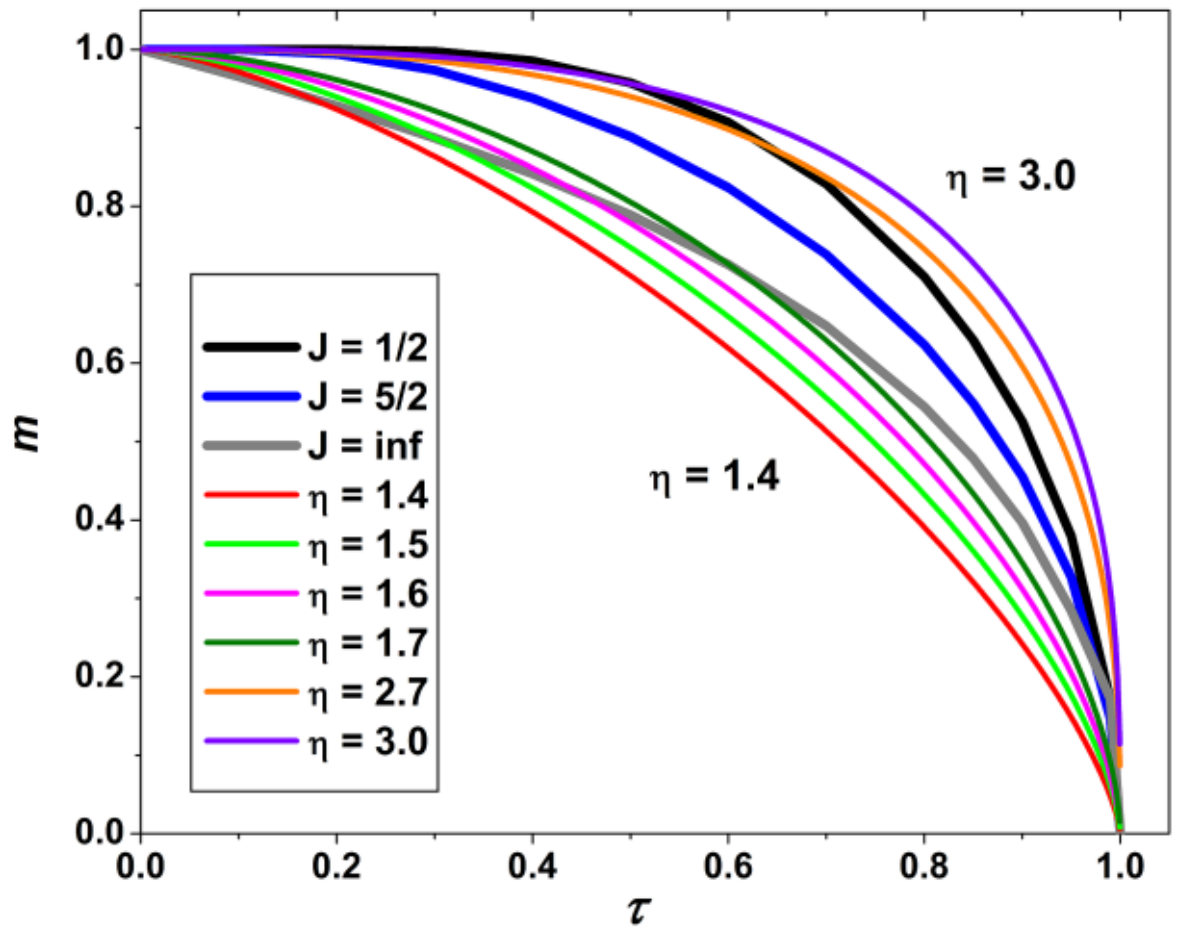

Fig. 11. Brilluin and Lamé curves.

## 5. Conclusions.

In this work we analyzed the shape of temperature dependence of spontaneous magnetization on about forty ferromagnetic materials. The shape parameter - squareness was determined from the curves fits by the superellipse equation (Lamé curve). It is similar to the quantum number $J$ in the Brillouin theory, but in contrast to it the superellipse equation covers all possible experimental curves. The agreement with experimental data was good for most materials including antiferromagnets. The squareness parameter (the power coefficient in the superellipse equation) is connected to the coupling strength between the nuclei vibrations and magnetic moments of electrons similar to the Curie temperature. It was in the range from 1.4 for antiferromagnets to 3.0 for iron. The squareness parameter was studied as a function of the Curie temperature, $T_C$. Both of them depend on the coupling between atomic vibrations and magnetic moments of electron. So, they can't be independent, but they are not proportional to each other if we consider all magnetic materials. For metallic alloys the general tendency was observed – squareness increase with the Curie temperature increase. The only exception was cobalt that has the same magnetization curve in the reduced coordinates as nickel despite of two times higher Curie temperature. Adding to nickel either ferromagnetic or nonferromagnetic metals leads to decrease of the curve squareness. The exception is Ni-Co alloys, for which the magnetization curve in the reduced coordinates is the same for all compositions starting from pure nickel to pure cobalt. No influence of the thermal expansion coefficient on the magnetization curve was observed – the zero-expansion invar have a standard shape following the Lamé curve identical to the curve of $Ni_{58}Cu_{42}$.

Measurement of the spontaneous magnetization as a function of temperature is not an easy task, especially around the Curie temperature. Usually, only one parameter - the Curie temperature is reported and the magnetization curve at some fixed field. At this moment there is data only for a very small amount of all magnetic materials. Most measurements were done from 1960s to 1990s. In this work we created a database of spontaneous magnetization curves for about 40 materials based on data present in literature till now. To understand fully mechanisms of the coupling between atomic vibrations and the magnetic moments, this database has to be extended – spontaneous magnetization curves have be measured on other materials, for which now only the Curie temperature is known.

## Declaration of competing interest

The authors declare that they have no known competing financial interests or personal relationships that could have appeared to influence the work reported in this paper.

## Acknowledgments

The authors acknowledge the assistance provided by the Ferroic Multifunctionalities project, supported by the Ministry of Education, Youth, and Sports of the Czech Republic. Project No. CZ.02.01.01/00/22_008/0004591, co-funded by the European Union. The single crystal preparation was partially performed in MGML (http://mgml.eu/), which was also supported within the program of Czech Research Infrastructures (project no. LM2023065),

## Data availability

The data that support the findings of this study are available from the corresponding authors upon request. The data are also available in Zenodo at https://doi.org/ 10.5281/zenodo.20446309 (Ref. 38).